\documentclass[conference]{IEEEtran}
\IEEEoverridecommandlockouts
\usepackage{comment}
\usepackage{cite}
\usepackage{stfloats}
\usepackage{amsmath,amssymb,amsfonts}
\usepackage{algorithmic}
\usepackage{graphicx}
\usepackage{textcomp}
\usepackage{xcolor}
\usepackage{multirow}
\def\BibTeX{{\rm B\kern-.05em{\sc i\kern-.025em b}\kern-.08em
    T\kern-.1667em\lower.7ex\hbox{E}\kern-.125emX}}
\begin{document}

\title{Leveraging Graph Embeddings for Opinion Leader Detection
\thanks{
\IEEEauthorrefmark{1} Corresponding authors.

\IEEEauthorrefmark{2} These authors contributed equally to this paper.}
}

\author{
    \IEEEauthorblockN{Yunming Hui\IEEEauthorrefmark{1}\IEEEauthorrefmark{2}, Luuk Buijsman\IEEEauthorrefmark{2}, Mel Chekol, Shihan Wang\IEEEauthorrefmark{1}}
    \IEEEauthorblockA{Department of Information and Computing Sciences, Utrecht University, Utrecht, The Netherlands}
    \IEEEauthorblockA{y.hui@students.uu.nl, luukbuijsman@gmail.com, \{m.w.chekol, s.wang2\}@uu.nl}
}
\maketitle

\begin{abstract}
Nowadays, social media plays an important role in many fields, such as the promotion of measures against major infectious diseases, merchandising, etc. In social media, some people are known as opinion leaders due to their strong ability to influence the opinions of others. The detection of opinion leaders has become an important task in social network analysis. Social networks are often represented in the form of graphs which allows a large number of graph analysis
methods to be used for opinion leader detection. 
Some studies have attempted to apply graph representation learning for opinion leader detection and achieved good results.

In this paper, we propose a model-agnostic framework that formulate the opinion leader detection problem as a ranking task 
of node embeddings.
A variety of methods and datasets are chosen to analyze the performance of our framework both qualitatively and quantitatively. Based on the analysis results, we propose a strategy that combines opinion leaders detected by two different ranking algorithms to obtain a more comprehensive set of opinion leaders. And we analyze the temporal changes of the opinion leaders in one of the dynamic social networks.
\end{abstract}


\section{Introduction}
In social networks, some people who are connected closely and excel in certain areas (e.g., high social status, articulate) are more capable of influencing the opinions or attitudes of people in their networks. These people are known as opinion leaders\cite{li2011talking}. Because of their high influence, these individuals have a significant impact on mass events such as raising public awareness of public health events, guiding marketing trends, etc.\cite{katarya2019survey}. Therefore, it is valuable to detect opinion leaders and analyze the patterns and characteristics that exist among them.

Social networks are usually represented in the form of graphs\cite{wellman2008development}, which allows a large number of graph analysis methods to be used for opinion leader detection. Compared to most graph analysis strategies, graph embedding allows graph analysis to be performed efficiently 
by transforming the graph into a low-dimensional vector space that preserves graph information\cite{cai2018comprehensive}. Several recent studies have proposed methods to apply graph embedding to social network analysis and have shown great performance\cite{bamakan2019opinion,luo2019identification,bo2020social}. However, for the opinion leader detection problem, methods proposed by existing studies are limited to a single graph embedding method and cannot be generalized. 
In this paper, we model a given dataset as a (dynamic) network, formulate the problem of opinion leader detection as a ranking task in graph representation learning and leverage node embeddings for identifying opinion leaders. Moreover, we propose a framework that is flexible to be utilized with various graph embedding models and ranking algorithms, and apply it to detect opinion leaders in dynamic social networks.
The contributions of this paper are the following: (1) One advantage of our framework is that it is model agnostic, i.e., various graph embedding models and ranking algorithms can be combined for opinion leader detection. (2) We carry out both quantitative and qualitative analysis to assess the performance our framework on two real-world datasets and compare across three graph embedding methods with two ranking algorithms, (3) We propose a strategy to combine opinion leaders discovered by different combinations of graph embedding models and ranking algorithms. (4) To the best of our knowledge, this is the first work utilizing graph embedding methods for opinion leader detection in dynamic social networks. 

\section{RELATED WORK} \label{section_relatedwork}
\paragraph{Opinion leader detection} According to Bamakan et al. the current methods of opinion leader detection can be divided into six categories\cite{bamakan2019opinion}. The centrality metrics is one of the most common methods. The design of our framework also uses this idea.

Different centrality metrics measure the importance of a user under different aspects. Among them, topological centrality is especially important for opinion leader detection because it can express the position of the user in the network\cite{freeman1978centrality}. Risselada et al.\cite{risselada2016indicators} and Yang et al.\cite{yang2018identifying} empirically investigated the usability of degree centrality and closeness centrality in opinion leader detection. Chen et al.\cite{chen2012identifying} proposed a semi-local centrality measure to solve the drawback that centrality metrics such as degree centrality cannot be applied to large and complex networks.

A crucial point in opinion leader detection is that users' influence and prestige are both very important to its centrality. Influence is determined by who the user pays attention to and prestige is determined by who pays attention to the user\cite{bamakan2019opinion}. In graph analysis, it can be expressed that both the incoming and outgoing links of a node have a significant impact on the importance of a node. The most influential algorithm that incorporates this idea is PageRank\cite{page1999pagerank}. Therefore, there are a large number of opinion leader detection methods serving different networks and purposes based on PageRank \cite{bamakan2019opinion}, such as InfluenceRank\cite{song2007identifying}, TrustRank\cite{chen2014identifying}. For microblogging-based social networks such as Twitter, Twitterrank\cite{weng2010twitterrank} is one of the most influential algorithms. It is proposed based on \textit{reciprocity} that can be explained by the phenomenon of homogeneity. It takes into account both topic similarity and link structure among users when measuring influence. LeaderRank\cite{lu2011leaders} is another effective method proposed based on PageRank. It introduces ground node, which makes it adaptive and parameter-free.

Recently, several studies have applied graph embedding to detect opinion leaders. Yang et al.\cite{bamakan2019opinion} used DeepWalk graph embedding method to generate embedding vectors of users then combined with network topology information to propose a local centrality index of network nodes to identify high-impact nodes. Luo et al.\cite{luo2019identification} used SNE (Social Network Embedding) model to obtain embedding vectors for each user. These embedding vectors were used to calculate the structure and text similarity between network nodes to improve the PageRank algorithm. Bo et al.\cite{bo2020social} proposed a method based on node embedding to integrate many types of interaction into embedded vectors. Then they used the vectors to define a new closeness measure to quantify closeness between users and incorporated the closeness measure into the ranking mechanism to propose a new PageRank-based influence ranking algorithm.


These graph embedding-based opinion leader detection methods are limited to only one graph embedding method. However, different graph embedding methods retain different graph information due to the different strategies they use\cite{cai2018comprehensive}, so there are limitations to these methods. 
In this work, we propose a model agnostic framework that is flexible to be utilized with various graph embedding models. 

\paragraph{Graph embedding} Graph embedding methods can be classified into three categories: factorization-based, random walk-based, and deep learning-based. Factorization-based graph embedding algorithms are out of the scope of our discussion because they cannot preserve graph connectivity information\cite{goyal2018graph,qiu2018network}.

The first approach applying random walk to graph embedding methods is DeepWalk\cite{perozzi2014deepwalk}. DeepWalk treats the local information obtained by random walk as sentences and uses language modelling and unsupervised feature learning (or deep learning) to learn the underlying representations. In comparison with algorithms that are allowed to observe global information, DeepWalk still performs better, especially when information is missing, which means that DeepWalk requires fewer data. DeepWalk is an online algorithm and can be performed in parallel. These advantages have made DeepWalk widely used in various fields since its creation. node2vec\cite{grover2016node2vec} is similar to deepwalk, but node2vec can perform a biased random walk based on BFS and DFS sampling to make sure that both homophily and structural equivalences are taken into account. Based on deepwalk and node2vec researchers have made many extensions, such as HARP, Walklets, etc.

With the development of deep learning, researchers prefer to use various types of deep learning frameworks for graph embedding. In particular, deep autoencoders are widely used due to their ability to model nonlinear structures in data\cite{goyal2018graph}. The method proposed by Wang et al. \cite{wang2016structural} uses this idea to thereby capture the highly nonlinear network structure. Kipf et al.\cite{kipf2016semi} proposed that Graph convolutional networks (GCN) have superior performance on large sparse graphs. Attributed Social Network Embedding (ASNE) proposed by Liao et al. \cite{liao2018attributed} improves network embedding by adding attributes of nodes. They modeled structural proximity and attribute proximity using a deep neural network architecture. The final ASNE model integrates models of structure and attributes through early fusion of input layers.

To verify the generality of our framework, we applyed DeepWalk and node2vec based on random walk and ASNE based on deep learning to our framework and analyze the performance in Section \ref{section_exp}, respectively.

\section{METHODOLOGY}
In this section, we first introduce our framework. Then we describe the graph structures, graph embedding models and ranking algorithms used to demonstrate our framework.

\subsection{Overall Framework}
\begin{figure}[htbp]
\centerline{\includegraphics[width=0.48\textwidth]{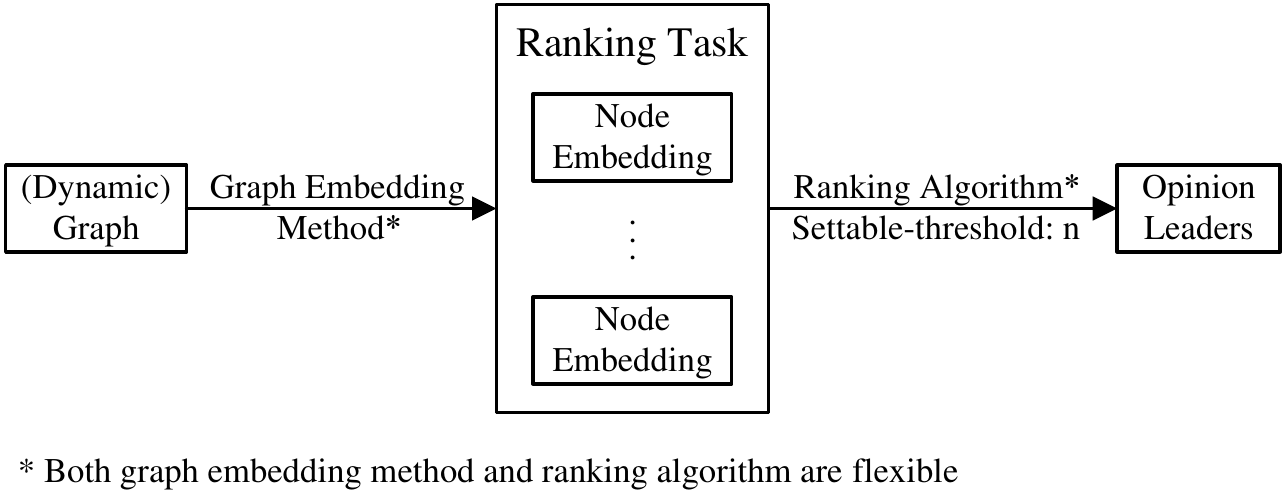}}
\caption{Graph embedding-based opinion leader detection framework}
\label{fig_framework}
\end{figure}

Our proposed framework is shown in Figure \ref{fig_framework}. In this framework, a (dynamic) graph is used to represent the (dynamic) interactions or relations among users in social networks.
First, the graph is transformed into a low-dimensional vector space using an graph embedding method so that each node is represented by an embedding vector. Then, the opinion leader detection task is transformed into a ranking task for users. Based on the generated embedding vectors, the users' influence is ranked using a
ranking algorithm. Finally, there is a settable threshold $n$, and the top $n$ users are considered as opinion leaders. In the framework, both embedding method and ranking algorithm are flexible.

\subsection{Methods}\label{Evaluation_Methodology}

\subsubsection{Graph representation} 


We represent a social network as a directed unweighted graph $G=\{V,E,f\}$, where $V$ is the node set, $f:V\rightarrow \mathcal{N}$ where $\mathcal{N}$ is the set of attributes which covers the attributes of all nodes, and $E$ is the edge set. In this work, when there is a directed edge pointing from $v_i$ to $v_j$, it means that user $i$ is a follower of user $j$, and user $j$ is a followee of user $i$.


\subsubsection{Graph Embedding Method}
We choose three representative graph embedding methods introduced in Section \ref{section_relatedwork}: DeepWalk\cite{perozzi2014deepwalk}, node2vec\cite{grover2016node2vec}, and ASNE\cite{liao2018attributed}.  
%
%
For all three methods, we strictly follow the approach proposed in the original papers. Both DeepWalk and node2vec only require the structural information of the graph as input. ASNE additionally requires each user's encoded attribute vector as input. The structure of used vectors is described in Section \ref{parameter_setting}.

\subsubsection{Ranking Algorithm}
Two ranking algorithms are also chosen to accomplish the ranking task and both of them
make good use of the results generated by graph embedding methods.
The first one is proposed by Yang et al.  \cite{yang2021identifying}. We refer to it as NLCRank in this paper. NLCRank features a combination of node topology information, which is calculated as follows:
\begin{equation}
\mathit{N L C} ( i ) = \sum _ { j \in \Gamma ( i ) } \mathit{Ks} _ { i } \times e ^ { - | \chi _ { i } - \chi _ { j } | ^ { 2 } }
\end{equation}
where $\mathit{Ks} _ { i }$ represents the K-shell value of node $i$, $\chi _ { i }$ and $\chi _ { j }$ are embedding vectors of user/node $i$ and $j$, and $\Gamma ( i )$ represents the third-level neighborhood of node $i$.

The second algorithm we employ is based on \cite{luo2019identification} and we refer to it as ASNERank. The original method can only be applied to graph embedding results from ASNE because of using intermediate results generated by ASNE. The intermediate results are used to calculate the confidence value of user $i$ is affected by user $j$.

To extend the generalization of this algorithm so that it can be applied to different graph embedding methods, we
recalculate the confidence value by computing the cosine similarity of the embedding vectors of two users. The specific equation is:
\begin{equation}
W _ { i , j } = \left\{ \begin{array} { c c } { \operatorname { exp } ( { u } _ { j } \cdot { u } _ { i } ) , } & { \text { if } j \text { is } i ^ { \prime } s \text { friend } } \\ { 0 } & { , \text { otherwise } } \end{array} \right.
\end{equation}
where ${ u } _ { j }$ and ${ u } _ { i }$ are the graph embedding results of user $j$ and user $i$, respectively.

Our rationale for this extension is as follows. The similarity between two embedding vectors represents the similarity between two nodes (users) \cite{arsov2019network}. And the more similar two users are, the more similar their opinions should be. Therefore the similarity of two embedding vectors can be used to measure the similarity of two users' opinions. Provided that one user is already following the other user, the higher the similarity of opinions is usually the more influenced by the other user is. In summary, the similarity of embedding vectors can be used to represent the confidence value.

\subsubsection{Analysis Methods} \label{subsection_Analysis}
In the course of our analysis, both qualitative and quantitative analysis are performed. For quantitative analysis, we evaluate our methods using the SIR model, which is widely used to assess the ability of opinion leaders to influence the networks\cite{rochert2022two}. The SIR model is originally used to simulate the spread of infectious diseases in a population. There are three types of nodes: susceptible, infected, and immune. To verify the ability of opinion leaders to influence the given network, we initially set the opinion leaders as infected, and then perform the simulation of infection procedures. After stabilization, the influence ability of the opinion leaders is evaluated by the final number of infected people in the network. The higher number of infections indicates the better performance of a method. 
%

We conducted qualitative analysis in terms of the user's identity, attitude towards keeping social distance, the number of follower and followee, etc. to validate and interpret the temporal changes of opinion leaders. 

\section{EXPERIMENTAL ANALYSIS} \label{section_exp}
In this section, we present our experimental evaluation. We use two datasets and three graph embedding models (DeepWalk, node2vec and ASNE) with two ranking methods (ASNERank and NLCRank). 
As a baseline, we use Twitterrank and LeaderRank. Since the Twitch dataset does not contain textual attributes making Twitterrank inapplicable, only LeaderRank is used as the baseline.  

\subsection{Dataset Description} 
% ...
The first dataset we used contains tweets written in Dutch which was collected by the twiqs.nl service \cite{sang2013dealing}. The used data contains tweets related to COVID-19 posted in 2020. Each tweet includes not only basic information such as content and publisher, but also the topic to which it belongs and the attitude towards this topic. The topics of tweet were obtained through the analysis of specific keywords, while the attitude is measured using three values: the probability of support, irrelevance and rejection, generated by using the classifiers trained on manually annotated data\cite{wang2020public}. 

In this paper, we used all tweets related to keeping social distance posted between the 40th and 44th week of 2020, a total of 56,173 tweets. In order to analyze dynamic changes of opinion leaders over time, we divided the data into five groups by week. We used the Twitter API\footnote{https://developer.twitter.com/en/products/twitter-api} to get the relationship (following/friendship relations) between all the users who posted these tweets, a total of 24,588 users. 
Since Twitter does not provide a user's history of friendships, we use the user's friendships at the time of our data collection query to generate the node links.
And as Twitter's friend relationships are directed, we follow the graph representation described in Section \ref{Evaluation_Methodology}. In particular, the data in every week is utilized to generate one graph, which corresponds to the snapshot of a big dynamic social network from this data. 
Here, each node is a Twitter user posted a tweet in that week, while each edge represents the following relation between two users (nodes).
The number of tweets per week and the size of each graph can be found in Table \ref{tab1}.

\begin{table}[htbp]
\caption{Number of tweets per week and size of the generated snapshot graph}
\begin{center}
\begin{tabular}{c c c c c c}
\hline
\textbf{Week} & \textbf{40} & \textbf{41} & \textbf{42} & \textbf{43} & \textbf{44} \\
\hline
\textbf{Tweets} & {9,604} & {10,922} & {14,766} & {9,091} & {7,311} \\
\textbf{Nodes} & {5,980} & {7,507} & {10,215} & {5,707} & {4,606} \\
\textbf{Edges} & {288,571} & {487,658} & {688,904} & {281,078} & {215,321} \\
\hline
\end{tabular}
\label{tab1}
\end{center}
\end{table}

The average number of Tweets sent per user during this period is 1.14. The mean and median of the number of all users' followers and followees are shown in Table \ref{tab2}.

\begin{table}[htbp]
\caption{Overall properties of users in each week}
\begin{center}
\begin{tabular}{c c c c c c}
\hline
\textbf{Week} & \textbf{40} & \textbf{41} & \textbf{42} & \textbf{43} & \textbf{44} \\
\hline
\textbf{Followers\_mean} & {340} & {468} & {468} & {315} & {317} \\
\textbf{Followers\_median} & {347} & {787} & {526} & {359} & {252} \\
\textbf{Followees\_mean} & {403} & {591} & {649} & {430} & {399} \\
\textbf{Followees\_median} & {426} & {918} & {746} & {507} & {453} \\
\hline
\end{tabular}
\label{tab2}
\end{center}
\end{table}

The second dataset we used contains several Twitch user-user networks \cite{rozemberczki2021multi}. All nodes in these networks have attributes, including the number of days and views, whether is mature and whether the user is a partner. In this paper, we use the network that contains users who speak French. 
We follow the definition in Section \ref{Evaluation_Methodology} to represent this static dataset as a graph. The generated graph contains 6,549 nodes and 112,666 edges.

\subsection{Parameter Setting} \label{parameter_setting}
For DeepWalk, the parameters are set as follows: walk length is 80 for Twitter dataset and 40 for Twitch dataset, embedding vector dimension is 64 for both, the number of walks per node is 10 for Twitter and 80 for Twitch dataset and the window size is both 10. The parameter settings of node2vec are the same as those of DeepWalk except that the dimension of embedding vector size of Twitter dataset is changed to 128. The parameters $p$ and $q$ used to control BFS (breadth-first search) and DFS (depth-first search) sampling are set to 0.25 and 4, respectively. 

\begin{figure}[htbp]
\centerline{\includegraphics[width=0.45\textwidth]{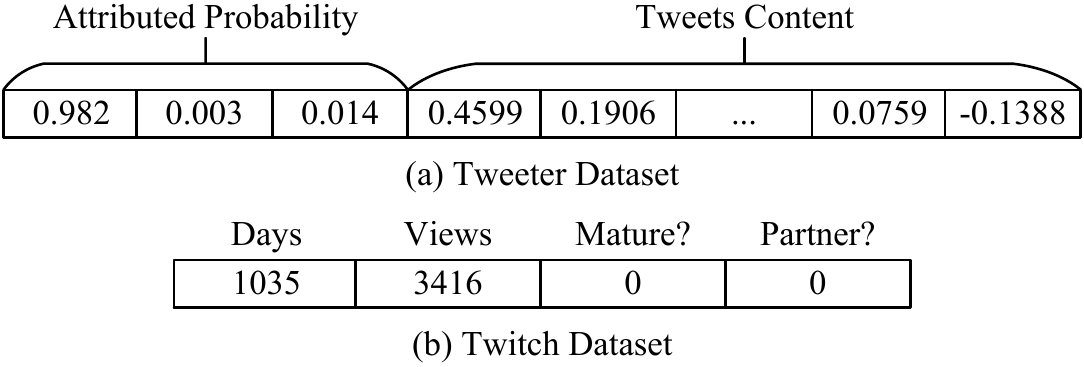}}
\caption{Two examples of input attribute vector of ASNE}
\label{fig_input_vector}
\end{figure}

For ASNE, the dimension of input attribute vector of the Twitter dataset is 13. As shown in Figure \ref{fig_input_vector}(a), the first 3 columns are user's attitude towards keeping social distance and the last 10 columns are generated by encoding the content of all tweets that each user published. Tweets are merged after removing nonsensical content such as stop words, special symbols, and network chains, and then we  encoded the merged tweets using a pre-trained language model named FastText\cite{grave2018learning}. The dimension of input attribute vector of the Twitch database is 4. As shown in Figure \ref{fig_input_vector}(b), are all the attributes of the user. 

The other parameters of ASNE are set as follows: the epoch for Twitter dataset is 20 and 30 for Twitch dataset, the batch size is 128, and the learning rate is 0.001. The embedding vector generated by ASNE is divided into two parts: structure and attributes. For the Twitter dataset, the dimensions of the structure and attribute parts are 20 and 40, respectively, and for the Twitch dataset, they are 60 and 40, respectively.

There are two important parameters in the SIR model, the first is the recovery rate per node ($\gamma$) and the second is the contagion rate per edge ($\tau$), which represents the ease of information propagation in the network. We set up two kinds of networks for easy and hard propagation validation.  In the \textit{easy-to-propagate} model, $\gamma$ is 1 and $\tau$ is 0.5, and in the \textit{hard-to-propagate} model, $\gamma$ is 1 and $\tau$ is 0.015. In the quantitative analysis using the SIR model, we set the settable threshold $n$ to be 100. To avoid the effect of the randomness of the simulations, the SIR validation experiments were repeated 50 times, and the SIR-related results shown next are the average of the 50 results.

\subsection{Evaluation \& Results}
In this subsection, we quantitatively evaluate the performance of \textit{six} methods using the framework on both two datasets. Also, qualitative evaluation is conducted on the unique Twitter dataset. To the best of our knowledge, no study has ever been done to analyze opinion leaders in such a dynamic social network.

\subsubsection{Quantitative Evaluation}
\begin{figure*}[htbp]
\centerline{\includegraphics[width=0.98\textwidth]{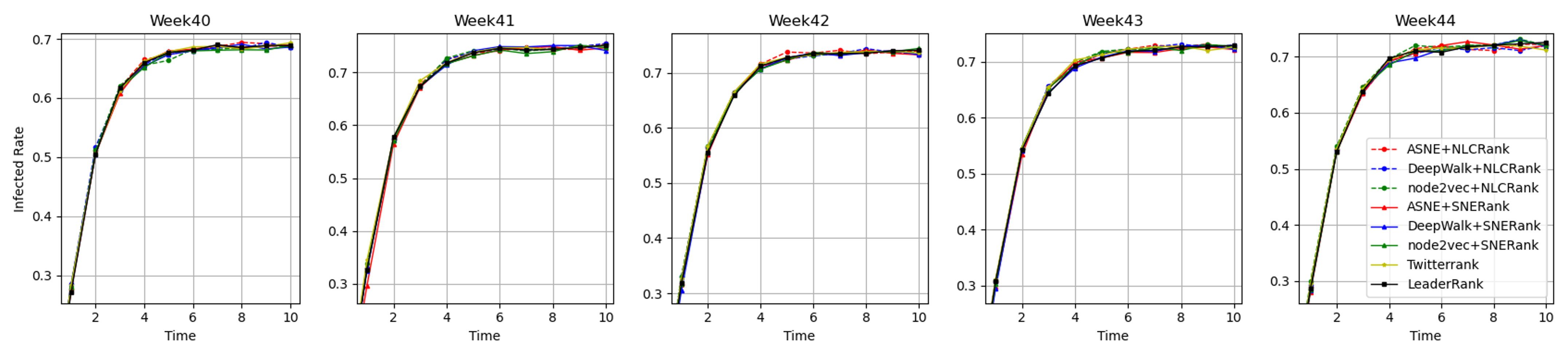}}
\caption{Validation results of six methods on Twitter dataset using easy-to-propagate SIR model}
\label{fig_SIR_easy}
\end{figure*}

\begin{figure*}[htbp]
\centerline{\includegraphics[width=0.98\textwidth]{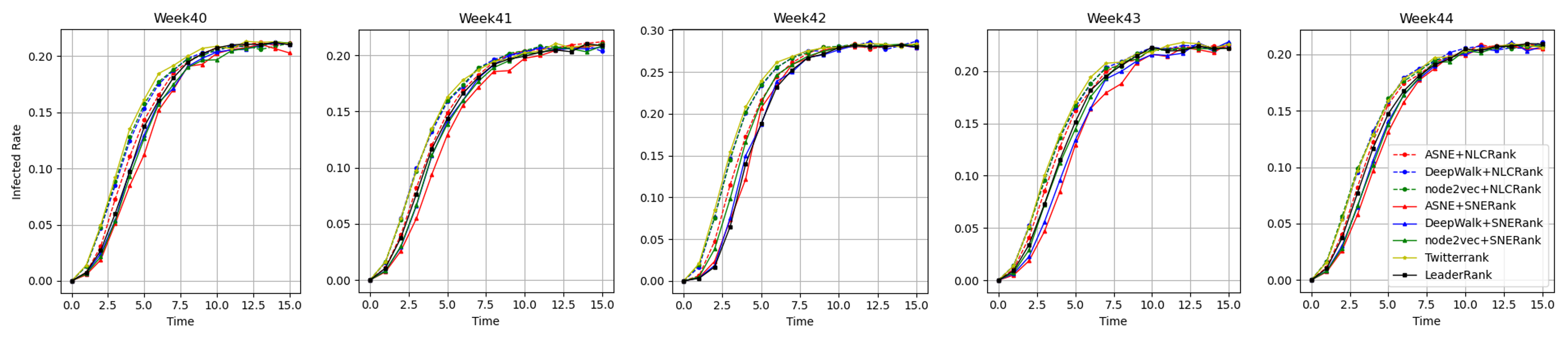}}
\caption{Validation results of six methods on Twitter dataset using hard-to-propagate SIR model}
\label{fig_SIR_hard}
\end{figure*}

\begin{figure}[htbp]
\centerline{\includegraphics[width=0.49\textwidth]{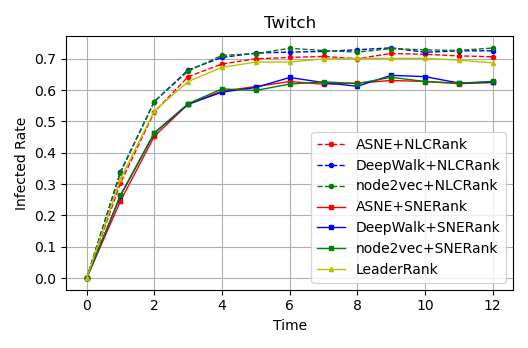}}
\caption{Validation results of six methods on Twitch dataset using hard-to-propagate SIR model}
\label{fig_SIR_twitch}
\end{figure}

We applied both easy-to-propagate and hard-to-propagate SIR models to quantitatively verify the performance of six methods on Twitter dataset. The results are shown in Figures \ref{fig_SIR_easy} and \ref{fig_SIR_hard}, respectively. Since the graph of Twitch dataset is sparser than that of Twitter dataset, we only used the hard-to-propagate model for verification, and the results are shown in Figure \ref{fig_SIR_twitch}.  

As can be seen from Figures \ref{fig_SIR_easy} and \ref{fig_SIR_hard}, the performance of all six methods is consistent from week to week, so no distinction between weeks will be made during the quantitative evaluation. On the easy-to-propagate SIR model, the difference in their performance is not significant, both in terms of the rate of transmission and the number of people eventually infected are similar. And, they all perform similarly to the baseline algorithms. From the results obtained using the hard-to-transmit SIR model, it can be seen that Twitterrank performs slightly better than the other six methods, with the difference being mainly in the speed of transmission and no significant difference in the number of final infections. 

On Twitch dataset, the results show that the three methods using NLCRank all perform slightly better than the LeaderRank. However, the performance of the three methods using ASNE is slightly worse than the baseline algorithm in terms of the speed of transmission and the number of final infections, but the differences are not significant.

The quantitative analysis results on both databases show that the performance of the methods using our proposed framework is similar to or even better than the baseline algorithms.

\subsubsection{Qualitative Evaluation}
To further valid our methods, we also conduct qualitative analysis to evaluate practical value of detected opinion leaders. In particular, we illustrate the practical implications of the opinion leaders detected by these methods. In qualitative evaluation, we set the settable-threshold $n$ to be 5 (i.e. find the topic 5 opinion leaders). The opinion leaders detected using the described method and set parameters are shown in Table \ref{tab_re_ASNERank} and \ref{tab_re_NLCRank}.

\begin{table*}[]
\caption{Top 5 opinion leaders detected by three methods using ASNERank}
\resizebox{\textwidth}{!}{%
\begin{tabular}{c|ccc|ccc|ccc|ccc|ccc}
\hline
WEEK                                                       & \multicolumn{3}{c|}{Week40}           & \multicolumn{3}{c|}{Week41}         & \multicolumn{3}{c|}{Week42}           & \multicolumn{3}{c|}{Week43}         & \multicolumn{3}{c}{Week44}         \\ \hline
\begin{tabular}[c]{@{}c@{}}Embedding\\ Method\end{tabular} & Name            & Follower & Followee & Name          & Follower & Followee & Name            & Follower & Followee & Name          & Follower & Followee & Name         & Follower & Followee \\ \hline
\multirow{5}{*}{ASNE}                                      & MinPres         & 2253     & 10       & MinPres       & 2568     & 2        & rivm            & 2532     & 51       & hugodejonge   & 1135     & 34       & nrc          & 913      & 2        \\
                                                           & hugodejonge     & 1343     & 39       & SBergsma      & 1461     & 31       & Nieuwsuur       & 2193     & 44       & vanranstmarc  & 1179     & 98       & op1npo       & 838      & 33       \\
                                                           & thierrybaudet   & 1485     & 4        & mauricedehond & 1891     & 21       & RTLnieuws       & 2329     & 38       & rivm          & 1493     & 34       & umarebru     & 821      & 93       \\
                                                           & MarijnissenL    & 976      & 37       & FleurAgemaPVV & 1359     & 16       & hugodejonge     & 1952     & 59       & RTLnieuws     & 1319     & 22       & RVeelo       & 742      & 26       \\
                                                           & claudiadebreij  & 1055     & 68       & RTLnieuws     & 1789     & 28       & volkskrant      & 2355     & 51       & telegraaf     & 1067     & 36       & wierdduk     & 1006     & 261      \\ \hline
\multirow{5}{*}{DeepWalk}                                  & MinPres         & 2253     & 10       & MinPres       & 2568     & 2        & volkskrant      & 2355     & 51       & rivm          & 1493     & 34       & EenVandaag   & 791      & 82       \\
                                                           & thierrybaudet   & 1485     & 4        & mauricedehond & 1891     & 21       & Nieuwsuur       & 2193     & 44       & FTM\_nl       & 1400     & 14       & wierdduk     & 1006     & 261      \\
                                                           & hugodejonge     & 1343     & 39       & hugodejonge   & 1521     & 37       & rivm            & 2532     & 51       & vanranstmarc  & 1179     & 98       & 1970\_R\_Mat & 331      & 429      \\
                                                           & telegraaf       & 1209     & 44       & vanranstmarc  & 1349     & 89       & RTLnieuws       & 2329     & 38       & bslagter      & 886      & 17       & op1npo       & 838      & 33       \\
                                                           & claudiadebreij  & 1055     & 68       & FTM\_nl       & 1924     & 12       & HaraldDoornbos  & 1502     & 48       & RTLnieuws     & 1319     & 22       & nrc          & 913      & 2        \\ \hline
\multirow{5}{*}{node2vec}                                  & MinPres         & 2253     & 10       & mauricedehond & 1891     & 21       & Gaia\_Universe  & 954      & 875      & hugodejonge   & 1135     & 34       & DiederikSmit & 372      & 23       \\
                                                           & thierrybaudet   & 1485     & 4        & MinPres       & 2568     & 2        & Percolator\_HNJ & 1127     & 182      & WybrenvanHaga & 932      & 47       & EenVandaag   & 791      & 82       \\
                                                           & JaapJansen      & 807      & 140      & RTLnieuws     & 1789     & 28       & telegraaf       & 1854     & 60       & ronald\_brok  & 418      & 431      & jndkgrf      & 787      & 180      \\
                                                           & woukevscherrenb & 908      & 66       & FTM\_nl       & 1924     & 12       & NOSsport        & 1120     & 4        & rivm          & 1493     & 34       & shossontwits & 847      & 94       \\
                                                           & NS\_online      & 995      & 127      & hugodejonge   & 1521     & 37       & rivm            & 2532     & 51       & RTLnieuws     & 1319     & 22       & RVeelo       & 742      & 26       \\ \hline
\end{tabular}%
}
\label{tab_re_ASNERank}
\end{table*}

\begin{table*}[]
\caption{Top 5 opinion leaders detected by three methods using NLCRank}
\resizebox{\textwidth}{!}{%
\begin{tabular}{c|ccc|ccc|ccc|ccc|ccc}
\hline
\multicolumn{1}{|c|}{WEEK}                                 & \multicolumn{3}{c|}{Week40}           & \multicolumn{3}{c|}{Week41}          & \multicolumn{3}{c|}{Week42}            & \multicolumn{3}{c|}{Week43}         & \multicolumn{3}{c}{Week44}            \\ \hline
\begin{tabular}[c]{@{}c@{}}Embedding\\ Method\end{tabular} & Name            & Follower & Followee & Name           & Follower & Followee & Name             & Follower & Followee & Name          & Follower & Followee & Name            & Follower & Followee \\ \hline
\multirow{5}{*}{ASNE}                                      & Kristine29ha    & 347      & 426      & rinushoogstad  & 385      & 695      & GielRuiter       & 221      & 217      & Ri\_qua0910   & 260      & 263      & Kristine29ha    & 326      & 396      \\
                                                           & SAVELnl         & 5        & 4        & Nonkelsjef     & 233      & 263      & Bloemkleur       & 220      & 105      & 2deKamerFVD   & 604      & 688      & EinarLinke      & 91       & 102      \\
                                                           & polprincip      & 186      & 189      & PeterK43579783 & 365      & 465      & VNeijs           & 1        & 5        & Kees71234     & 317      & 168      & HendrikAlbertII & 91       & 132      \\
                                                           & solitar66558765 & 212      & 203      & FricoGod       & 482      & 494      & jusraaijmakers   & 1        & 4        & GoThorium     & 323      & 535      & vm19630418      & 163      & 216      \\
                                                           & Volger66082047  & 237      & 235      & VeenInfo       & 625      & 698      & Nationalist\_NLD & 231      & 211      & huigen2       & 359      & 507      & jacnaber        & 170      & 142      \\ \hline
\multirow{5}{*}{DeepWalk}                                  & Kristine29ha    & 347      & 426      & FlapFriesland  & 787      & 918      & spill2012        & 526      & 746      & huigen2       & 359      & 507      & RolfvanZutphen  & 252      & 453      \\
                                                           & huigen2         & 373      & 511      & aslo63         & 310      & 339      & liekcornelissen  & 460      & 604      & FlapFriesland & 538      & 637      & spill2012       & 317      & 453      \\
                                                           & Wallyboytoaogm1 & 261      & 371      & angelb52       & 518      & 485      & Wegaandiep       & 567      & 775      & rinushoogstad & 253      & 472      & huigen2         & 335      & 456      \\
                                                           & AmberBrouwer2   & 256      & 357      & Wiep13396680   & 620      & 658      & marcvdhoogen     & 530      & 659      & esther241101  & 259      & 419      & MediaservicesEU & 338      & 385      \\
                                                           & MiekeHoogvliet  & 265      & 399      & spill2012      & 492      & 703      & FrankBrecht      & 616      & 679      & malavita666   & 299      & 535      & ang\_haar       & 249      & 247      \\ \hline
\multirow{5}{*}{node2vec}                                  & huigen2         & 373      & 511      & NassauWillem   & 544      & 783      & huigen2          & 544      & 771      & huigen2       & 359      & 507      & huigen2         & 335      & 456      \\
                                                           & Aspirides78     & 402      & 526      & angelb52       & 518      & 485      & AG28071973       & 423      & 718      & FlapFriesland & 538      & 637      & spill2012       & 317      & 453      \\
                                                           & Kristine29ha    & 347      & 426      & Medusa99160583 & 571      & 785      & spill2012        & 526      & 746      & malavita666   & 299      & 535      & Medusa99160583  & 364      & 515      \\
                                                           & 2tijro          & 207      & 233      & FlapFriesland  & 787      & 918      & HansDek35828687  & 695      & 785      & spill2012     & 337      & 470      & ducom99         & 538      & 671      \\
                                                           & colourbird00    & 422      & 502      & spill2012      & 492      & 703      & malavita666      & 464      & 846      & VeltmanA1     & 213      & 296      & RolfvanZutphen  & 252      & 453      \\ \hline
\end{tabular}%
}
\label{tab_re_NLCRank}
\end{table*}

Based on the results of the quantitative analysis, we divided the results into two groups for qualitative analysis according to the ranking algorithm used.

The opinion leaders detected by methods using ASNERank have two significant features. First, they all have a very large number of followers and a very small number of followee. The number of followers of all of them exceeded the median of the number of followers as listed in Table \ref{tab2}. Another feature is that the vast majority of these users are well-known Dutch public figures or run by large Dutch institutions. For example, "MinPres" is the Dutch Prime Minister Mark Rutte, "hugodejonge" is the Dutch Minister for Housing and Spatial Planning, "rivm" is operated by the Dutch National Institute for Public Health and the Environment", "telegraaf" is operated by the largest daily newspaper in the Netherlands, "jndkgrf" is the famous Dutch columnist Jan Dijkgraaf, and "woukevscherrenb" is the famous Dutch host Wouke van Scherrenburg.

This is reasonable because these people or organizations have strong credibility, which makes their opinions more acceptable to their followers. Moreover, these people or institutions are usually the ones who present their opinions in online social networks which makes them not need a large number of followees. In addition, national institutions such as the Dutch National Institute for Public Health  and the Environment ("rivm") are particularly influential, in our study on the topic of social distancing, because of the legally mandatory nature of their opinions.

Before discussing methods using NLCRank, there are some outliers that need to be accounted for first. These outliers mainly occur in the opinion leaders detected by the method with ASNE as the graph embedding model. For example, users "SAVELnl", "VNeijs" and "jusraaijmakers" have a very low number of followers and followees, and further analysis of their specific tweets and other information does not show any evidence that they should be opinion leaders. We believe that the main reason is  the authors of NLCRank use DeepWalk as the graph embedding method, and there is an order of magnitude difference between the modulus of the embedding vector generated by ASNE and the embedding vector generated by DeepWalk. Therefore, for ASNERank, we should pay attention to this issue when using non-DeepWalk family graph embedding methods.

Out of these abnormal users, the distinctive feature of the opinion leaders detected by methods using NLCRank is that the number of their followers and followees is relatively high, basically exceeding or approaching the average of all users listed in Table \ref{tab2}. In addition, unlike the previously mentioned opinion leaders, they are ordinary people. For example, according to their Twitter profiles, "huigen2" is a retired facility manager, "GielRuiter" is a volunteer, etc.

The plausibility of these opinion leaders can be explained using the view of Katz et al.\cite{katz2017personal}. Katz et al. argue that opinion leaders act as a medium between the mass media and the audience, they understand and analyze other people's views and ideas, and then make their analyzed views and ideas available to people. And these opinion leaders should not be well-known figures. The opinion leaders described by Katz et al. are very symbolic of the characteristics of the opinion leaders detected by methods using NLCRank. First of all they are indeed ordinary people. Secondly, since they need to understand and analyze others' views, they need to follow a large number of users to ensure the intake of a sufficient amount of information, and they need to have a sufficient number of followees to spread their opinions. 

The above analysis illustrates that our framework has good compatibility with both two datasets, three graph embedding methods and two ranking algorithms. This shows the potential of our framework that it can support various graph embedding methods and ranking algorithms. 

\begin{table}[htbp]
\caption{Attitudes toward "keeping social distance" among opinion leaders obtained by ASNERank and NLCRank and overall users}
\begin{center}
\begin{tabular}{ccccc}
\hline
Week                & Group    & Support & Reject & Irrelevant \\ \hline
\multirow{3}{*}{40} & ASNERank & 0.71    & 0.25   & 0.02       \\
                    & NLCRank  & 0.50    & 0.47   & 0.02       \\
                    & Overall users  & 0.53    & 0.44   & 0.02       \\ \hline
\multirow{3}{*}{41} & ASNERank & 0.82    & 0.13   & 0.03       \\
                    & NLCRank  & 0.29    & 0.65   & 0.05       \\
                    & Overall users  & 0.39    & 0.56   & 0.03       \\ \hline
\multirow{3}{*}{42} & ASNERank & 0.85    & 0.02   & 0.12       \\
                    & NLCRank  & 0.64    & 0.33   & 0.01       \\
                    & Overall users  & 0.65    & 0.30   & 0.04       \\ \hline
\multirow{3}{*}{43} & ASNERank & 0.77    & 0.16   & 0.05       \\
                    & NLCRank  & 0.54    & 0.35   & 0.08       \\
                    & Overall users  & 0.54    & 0.38   & 0.10       \\ \hline
\multirow{3}{*}{44} & ASNERank & 0.73    & 0.23   & 0.03       \\
                    & NLCRank  & 0.51    & 0.47   & 0.01       \\
                    & Overall users  & 0.53    & 0.45   & 0.05       \\ \hline
\end{tabular}
\label{tab3}
\end{center}
\end{table}

\begin{table*}[t]
\caption{final top 15 opinion leaders obtained by the combination strategy}
\begin{center}
\begin{tabular}{cccccc}
\hline
\multicolumn{1}{l}{}                                                     
& Week40          & Week41        & Week42          & Week43        & Week44         \\ \hline
\multirow{5}{*}{\begin{tabular}[c]{@{}c@{}}ASNERank\\ Part\end{tabular}} 
& MinPres         & MinPres       & rivm            & hugodejonge   & nrc            \\
& hugodejonge     & mauricedehond & Nieuwsuur       & rivm          & EenVandaag     \\
& thierrybaudet   & Sbergsma      & RTLnieuws       & WybrenvanHaga & DiederikSmit   \\
& telegraaf       & RTLnieuws     & VNeijs          & vanranstmarc  & op1npo         \\
& JaapJansen      & hugodejonge   & Percolator\_HNJ & FTM\_nl       & wierdduk       \\ \hline
\multirow{10}{*}{\begin{tabular}[c]{@{}c@{}}NLCRank\\ Part\end{tabular}}
& Kristine29ha    & rinushoogstad  & GielRuiter      & Ri\_qua0910   & Kristine29ha    \\
& huigen2         & FlapFriesland  & spill2012       & huigen2       & RolfvanZutphen  \\
& polprincip      & NassauWillem   & huigen2         & FlapFriesland & huigen2         \\
& Aspirides78     & angelb52       & liekcornelissen & 2deKamerFVD   & spill2012       \\
& Wallyboytoaogm1 & aslo63         & AG28071973      & malavita666   & ducom99         \\
& AmberBrouwer2   & Nonkelsjef     & Kristine29ha    & Kees71234     & vm19630418      \\
& MiekeHoogvliet  & PeterK43579783 & WimPekel        & rinushoogstad & MediaservicesEU \\
& 2tijro          & Wiep13396680   & Wegaandiep      & VeltmanA1     & Medusa99160583  \\
& Volger66082047  & Medusa99160583 & HansDek35828687 & esther241101  & jacnaber        \\
& kaatje1919      & huigen2        & marcvdhoogen    & GoThorium     & mislukt12345    \\ \hline
\end{tabular}
\label{tab4}
\end{center}
\end{table*}

\subsection{Analysis \& Interpretation}
In the qualitative evaluation in the previous subsection, we have made a valuable discovery: two types of opinion leaders are detected in our database through various methods. The first category are those public figures or well-known organizations with a large number of followers, detected by methods using ASNERank. The second type are the general public who act as intermediaries in the dissemination of opinions, detected by methods using NLCRank. It is clear that a comprehensive set of opinion leaders should include both types of opinion leaders. So, combining the results obtained by these two methods is a good strategy. Next, attitudes of the users will be analyzed to further illustrate why and how they should be combined. We calculated the evaluation attitudes of the weekly opinion leaders detected by methods using ASNERank and NLCRank, and users in the whole network, which are shown in Table \ref{tab3}. The "support" values of each week are calculated as follows:

\begin{equation}
\begin{aligned}
&support_\mathit{ASNERank} = \frac{\sum_{\substack{v\in V_\mathit{ASNERank}}} support_v} {|V_\mathit{ASNERank}|}\\
&support_\mathit{NLCRank} = \frac{\sum_{\substack{v\in V_\mathit{NLCRank}}} support_v} {|V_\mathit{NLCRank}|}\\
&support_\mathit{Overall} = \frac{\sum_{\substack{v \in V}} support_v} {|V|}
\end{aligned}
\end{equation}
where $V_\mathit{ASNERank}$ and $V_\mathit{NLCRank}$ are the sets of users who are detected as opinion leaders by methods using ASNERank and NLCRank respectively, $V$ is the set of all users. The values of "Reject" and "Irrelevant" are calculated similarly.

The attitude of a comprehensive group of opinion leaders should be consistent with that of the overall users. However, opinion leaders detected by methods using ASNERank have more positive attitudes towards keeping social distance and are significantly different from the overall users. By contrast, the opinion leaders detected by using the NLCRank method are more negative and closer to the overall users, but there is still a certain gap between them and the overall users. This is further evidence that the opinion leaders obtained by both types of algorithms are incomplete. In addition, the overall users' attitudes are between the attitudes of the two types of opinion leaders, so combining the opinion leaders obtained by both methods is a reasonable strategy.

Combining all of the above experimental results and findings, we propose a strategy that integrates the results of various methods. The basic idea is to combine the results of the methods with the same ranking algorithm equally. Since different graph embedding methods retain different graph information, the results of the methods using the same ranking algorithm should be combined in a way that takes into account the various graph embedding methods. It is worth noting that those outliers generated by methods that use ASNE and ASNERank should be excluded. Then, for the final opinion leaders, the proportion of opinion leaders detected by methods using NLCRank should be greater, as it can be seen in Table \ref{tab3}  that they are closer to the overall users' attitudes (here we choose a ratio of one to two).

\subsection{Dynamic Analysis}
The final top 15 opinion leaders detected according to the proposed strategy are shown in Table \ref{tab4}. We dynamically analyzed the characteristics of these opinion leaders over time and had the following findings.

First, the influence of a user in a social network does not change in the short term. Users such as "huigen2" and "hugodejonge" (although "hugodejonge" does not appear in Table \ref{tab4}, he is still ranked  in the top 50 in other weeks) have a high influence every week. Users like "MinPres" was selected as an opinion leader only in weeks 40 and 41, but the reason was that he did not post tweets related to keeping social distance in the remaining three weeks. According to our validation, the vast majority of users who appear in Table \ref{tab4} are ranked in the top 100, as long as they posted a tweet in any given week.

Second, the first category of opinion leaders with high credibility and a large number of followers are more likely to inject new and different views into social networks. As shown in Table \ref{tab3}, from week 40 to week 41, users in social networks became significantly more resistant to maintaining social distance. The second category of opinion leaders also became more negative. But on the contrary, the first category of opinion leaders' support for this issue is increasing, which makes the social network more positive in week 42.

\section{CONCLUSION AND FUTURE WORK}
Opinion leader detection is a very important and valuable issue in social network analysis. By using graph embedding, this problem can be solved more easily and efficiently. We formulate the opinion leader detection problem as a ranking task in graphical representation learning, and propose a graph embedding-based opinion leader detection framework that is compatible with any intent embedding method. By validation, our framework has good performance on different datasets using different graph embedding methods and ranking algorithms. In addition, we propose a strategy to integrate the detection results of different methods 
And we analyzed the characteristics of the change of opinion leaders over time.

We only studied the opinion leaders in the snapshots of dynamic social networks in this paper, for the future work, we plan to integrate the dynamic features of a network in the graph embedding step to improve the opinion leader detection in dynamic social networks.



\bibliographystyle{ieeetr}
\bibliography{ref}

\begin{thebibliography}{10}

\bibitem{li2011talking}
F.~Li and T.~C. Du, ``Who is talking? an ontology-based opinion leader
  identification framework for word-of-mouth marketing in online social
  blogs,'' {\em Decision support systems}, vol.~51, no.~1, pp.~190--197, 2011.

\bibitem{katarya2019survey}
R.~Katarya and D.~Gautam, ``Survey on opinion leader in social network using
  data mining,'' in {\em 2019 5th International Conference on Advanced
  Computing \& Communication Systems (ICACCS)}, pp.~506--509, IEEE, 2019.

\bibitem{wellman2008development}
B.~Wellman, ``The development of social network analysis: A study in the
  sociology of science,'' {\em Contemporary Sociology}, vol.~37, no.~3, p.~221,
  2008.

\bibitem{cai2018comprehensive}
H.~Cai, V.~W. Zheng, and K.~C.-C. Chang, ``A comprehensive survey of graph
  embedding: Problems, techniques, and applications,'' {\em IEEE Transactions
  on Knowledge and Data Engineering}, vol.~30, no.~9, pp.~1616--1637, 2018.

\bibitem{bamakan2019opinion}
S.~M.~H. Bamakan, I.~Nurgaliev, and Q.~Qu, ``Opinion leader detection: A
  methodological review,'' {\em Expert Systems with Applications}, vol.~115,
  pp.~200--222, 2019.

\bibitem{luo2019identification}
J.~Luo, Y.~Du, R.~Li, and F.~Cheng, ``Identification of opinion leaders by
  using social network embedding,'' in {\em 2019 IEEE 5th International
  Conference on Computer and Communications (ICCC)}, pp.~1412--1416, IEEE,
  2019.

\bibitem{bo2020social}
H.~Bo, R.~McConville, J.~Hong, and W.~Liu, ``Social network influence ranking
  via embedding network interactions for user recommendation,'' in {\em
  Companion Proceedings of the Web Conference 2020}, pp.~379--384, 2020.

\bibitem{freeman1978centrality}
L.~C. Freeman, ``Centrality in social networks conceptual clarification,'' {\em
  Social networks}, vol.~1, no.~3, pp.~215--239, 1978.

\bibitem{risselada2016indicators}
H.~Risselada, P.~C. Verhoef, and T.~H. Bijmolt, ``Indicators of opinion
  leadership in customer networks: self-reports and degree centrality,'' {\em
  Marketing Letters}, vol.~27, no.~3, pp.~449--460, 2016.

\bibitem{yang2018identifying}
L.~Yang, Y.~Qiao, Z.~Liu, J.~Ma, and X.~Li, ``Identifying opinion leader nodes
  in online social networks with a new closeness evaluation algorithm,'' {\em
  Soft Computing}, vol.~22, no.~2, pp.~453--464, 2018.

\bibitem{chen2012identifying}
D.~Chen, L.~L{\"u}, M.-S. Shang, Y.-C. Zhang, and T.~Zhou, ``Identifying
  influential nodes in complex networks,'' {\em Physica a: Statistical
  mechanics and its applications}, vol.~391, no.~4, pp.~1777--1787, 2012.

\bibitem{page1999pagerank}
L.~Page, S.~Brin, R.~Motwani, and T.~Winograd, ``The pagerank citation ranking:
  Bringing order to the web.,'' tech. rep., Stanford InfoLab, 1999.

\bibitem{song2007identifying}
X.~Song, Y.~Chi, K.~Hino, and B.~Tseng, ``Identifying opinion leaders in the
  blogosphere,'' in {\em Proceedings of the sixteenth ACM conference on
  Conference on information and knowledge management}, pp.~971--974, 2007.

\bibitem{chen2014identifying}
Y.~Chen, X.~Wang, B.~Tang, R.~Xu, B.~Yuan, X.~Xiang, and J.~Bu, ``Identifying
  opinion leaders from online comments,'' in {\em Chinese national conference
  on social media processing}, pp.~231--239, Springer, 2014.

\bibitem{weng2010twitterrank}
J.~Weng, E.-P. Lim, J.~Jiang, and Q.~He, ``Twitterrank: finding topic-sensitive
  influential twitterers,'' in {\em Proceedings of the third ACM international
  conference on Web search and data mining}, pp.~261--270, 2010.

\bibitem{lu2011leaders}
L.~L{\"u}, Y.-C. Zhang, C.~H. Yeung, and T.~Zhou, ``Leaders in social networks,
  the delicious case,'' {\em PloS one}, vol.~6, no.~6, p.~e21202, 2011.

\bibitem{goyal2018graph}
P.~Goyal and E.~Ferrara, ``Graph embedding techniques, applications, and
  performance: A survey,'' {\em Knowledge-Based Systems}, vol.~151, pp.~78--94,
  2018.

\bibitem{qiu2018network}
J.~Qiu, Y.~Dong, H.~Ma, J.~Li, K.~Wang, and J.~Tang, ``Network embedding as
  matrix factorization: Unifying deepwalk, line, pte, and node2vec,'' in {\em
  Proceedings of the eleventh ACM international conference on web search and
  data mining}, pp.~459--467, 2018.

\bibitem{perozzi2014deepwalk}
B.~Perozzi, R.~Al-Rfou, and S.~Skiena, ``Deepwalk: Online learning of social
  representations,'' in {\em Proceedings of the 20th ACM SIGKDD international
  conference on Knowledge discovery and data mining}, pp.~701--710, 2014.

\bibitem{grover2016node2vec}
A.~Grover and J.~Leskovec, ``node2vec: Scalable feature learning for
  networks,'' in {\em Proceedings of the 22nd ACM SIGKDD international
  conference on Knowledge discovery and data mining}, pp.~855--864, 2016.

\bibitem{wang2016structural}
D.~Wang, P.~Cui, and W.~Zhu, ``Structural deep network embedding,'' in {\em
  Proceedings of the 22nd ACM SIGKDD international conference on Knowledge
  discovery and data mining}, pp.~1225--1234, 2016.

\bibitem{kipf2016semi}
T.~N. Kipf and M.~Welling, ``Semi-supervised classification with graph
  convolutional networks,'' {\em arXiv preprint arXiv:1609.02907}, 2016.

\bibitem{liao2018attributed}
L.~Liao, X.~He, H.~Zhang, and T.-S. Chua, ``Attributed social network
  embedding,'' {\em IEEE Transactions on Knowledge and Data Engineering},
  vol.~30, no.~12, pp.~2257--2270, 2018.

\bibitem{yang2021identifying}
X.-H. Yang, Z.~Xiong, F.~Ma, X.~Chen, Z.~Ruan, P.~Jiang, and X.~Xu,
  ``Identifying influential spreaders in complex networks based on network
  embedding and node local centrality,'' {\em Physica A: Statistical Mechanics
  and its Applications}, vol.~573, p.~125971, 2021.

\bibitem{arsov2019network}
N.~Arsov and G.~Mirceva, ``Network embedding: An overview,'' {\em arXiv
  preprint arXiv:1911.11726}, 2019.

\bibitem{rochert2022two}
D.~R{\"o}chert, M.~Cargnino, and G.~Neubaum, ``Two sides of the same leader: an
  agent-based model to analyze the effect of ambivalent opinion leaders in
  social networks,'' {\em Journal of computational social science}, pp.~1--47,
  2022.

\bibitem{sang2013dealing}
E.~T.~K. Sang and A.~Van~den Bosch, ``Dealing with big data: The case of
  twitter,'' {\em Computational Linguistics in the Netherlands Journal},
  vol.~3, pp.~121--134, 2013.

\bibitem{wang2020public}
S.~Wang, M.~Schraagen, E.~Tjong Kim~Sang, and M.~Dastani, ``Public sentiment on
  governmental covid-19 measures in dutch social media,'' in {\em Proceedings
  of the 1st Workshop on NLP for COVID-19 (Part 2) at the 2020 Conference on
  Empirical Methods in Natural Language Processing}, Association for
  Computational Linguistics (ACL), 2020.

\bibitem{rozemberczki2021multi}
B.~Rozemberczki, C.~Allen, and R.~Sarkar, ``Multi-scale attributed node
  embedding,'' {\em Journal of Complex Networks}, vol.~9, no.~2, p.~cnab014,
  2021.

\bibitem{grave2018learning}
E.~Grave, P.~Bojanowski, P.~Gupta, A.~Joulin, and T.~Mikolov, ``Learning word
  vectors for 157 languages,'' {\em arXiv preprint arXiv:1802.06893}, 2018.

\bibitem{katz2017personal}
E.~Katz and P.~F. Lazarsfeld, {\em Personal influence: The part played by
  people in the flow of mass communications}.
\newblock Routledge, 2017.

\end{thebibliography}

\end{document}